# Pressure impact on the stability and distortion of the crystal structure of CeScO$_3$

Daniel Errandonea[1,*], David Santamaria-Perez[1], Domingo Martinez-Garcia[1], Oscar Gomis[2], Rakesh Shukla[3], S. Nagabhusan Achary[3], Avesh K. Tyagi[3], and Catalin Popescu[4]

[1]*Departamento de Física Aplicada-ICMUV, Universidad de Valencia, MALTA Consolider Team, Edificio de Investigación, C/Dr. Moliner 50, 46100 Burjassot, Valencia, Spain*

[2]*Centro de Tecnologías Físicas, MALTA Consolider Team, Universitat Politècnica de València, 46022 Valencia, Spain*

[3]*Chemistry Division, Bhabha Atomic Research Centre, Trombay, Mumbai 400085, India*

[4]*CELLS-ALBA Synchrotron Light Facility, Cerdanyola, 08290 Barcelona, Spain*

**Abstract:** The effects of high-pressure on the crystal structure of orthorhombic (Pnma) perovskite-type cerium scandate have been studied *in situ* under high pressure by means of synchrotron x-ray powder diffraction, using a diamond-anvil cell. We have found that the perovskite-type crystal structure remains stable up to 40 GPa, the highest pressure reached in the experiments. The evolution of unit-cell parameters with pressure has indicated an anisotropic compression. The room-temperature pressure-volume equation of state obtained from the experiments indicated the EOS parameters $V_0 = 262.5(3)$ Å$^3$, $B_0 = 165(7)$ GPa, and $B_0' = 6.3(5)$. From the evolution of microscopic structural parameters like bond distances and coordination polyhedra of cerium and scandium, the macroscopic behavior of $CeScO_3$ under compression has been explained and reasoned for its large pressure stability. The reported results are discussed in comparison with high-pressure results from other perovskites.

I.   Introduction

Perovskite-structured materials continue to attract the attention of scientists because of their remarkable properties, which include colossal magneto-resistance, metal-insulator transitions, piezoelectricity, ferroelectricity, etc. These properties make perovskites candidates for various technological applications.[1] Besides, the perovskite structure adopted by minerals, in particular $(Mg,Fe)SiO_3$, bear significant geological importance due to their formation in lower mantle of Earth.[2] The $ABO_3$ perovskite-type materials with both A and B being trivalent cations form the largest class of perovskites due to the wide choice of cations satisfying the tolerance criteria of the structure. Most of the rare-earth ions form a perovskite-type structure with trivalent transition metal ions and $Ga^{3+}$ and $Al^{3+}$. The crystal structure and structural transitions in such materials are related to their fundamental properties and thus they have a rich crystal chemistry. As a consequence of these facts, perovskites have been largely studied under high pressure (HP) or high temperature to understand their structural and physical properties as well as for searching new phases. Depending on the ionic radii of cations as well as external parameters like, pressure or temperature, the ideal perovskites distort and form different low-symmetry structures. It is commonly observed that the low-symmetry structures transform to high-symmetry structures with increasing temperature. However, the effect of pressure on perovskites is not well understood due to fact that transitions from a low-symmetry to a high-symmetry structure as well as from a high-symmetry to a low-symmetry structure are observed under compression. Thus diversified observations on phase transitions in perovskite-type compounds have been reported in literature.[2-7] As a case, the orthorhombic (Pnma) $LaGaO_3$ transforms to a rhombohedral ($R\bar{3}c$) structure at 150°C as well as at 2.5 GPa.[7, 8] On the contrary, the cubic perovskite-type fluorides $BaLiF_3$ and $KMgF_3$ remain stable up to significantly higher pressure, viz > 40 GPa.[9, 10]

On the other hand, although a number of structural studies on orthorhombic perovskites have been reported in the literature, the structural transitions are not unique. For instance, GdMnO$_3$ undergoes a isosymmetric structural transition around 50 GPa[11], LaGaO$_3$ and LaAlO$_3$ an orthorhombic-to-rhombohedral transition,[7, 12, 13] PbRuO$_3$ suffers an orthorhombic-orthorhombic transition (Pbnm to Pbn2$_1$)[4], and NaMgF$_3$ a different orthorhombic-orthorhombic transition.[14, 15] The HP transitions in perovskites have thus attracted significant interest due to the impact of pressure in their crystal chemistry, owing the influence of compression on the internal structural parameters, bond strength, and polyhedral properties, which play a dominating role in determining the properties of perovskites. In particular, Zhao *et al.*[6] explained that the structural stabilities of perovskites are related to the compressibility and distortion of AO$_{12}$ and BO$_6$ polyhedra and the compressibility of A-O and B-O bonds. The evolutions of atomic coordinates are thus important to understand the phenomena under pressure.

Although, varieties of perovskite-type oxides have been extensively studied under HP,[3 - 7, 11 – 13, 16 – 19] to the best of our knowledge nothing is known on the effect of pressure on rare-earth scandates. These scandates have an orthorhombic structure, where no pressure-induced electronic contribution, like overlapping of orbitals, is expected to affect their HP behavior. Therefore, the distortion and compressibility of bonds can be assumed to be the driving force of the structural modification under compression, being the pressure dependent structural behavior and lattice distortion easily interpreted. In addition, rare-earth scandates have technological relevance and are promising candidates for the replacement of silicon dioxide in metal-oxide-semiconductor field-effect transistors due to their large dielectric constant and optical band gap.[20] Among rare-earth scandates, CeScO$_3$ is an interesting member which can be easily produced by a facile chemical method.[21] Its thermal stability and electronic properties have been discussed.[21]

$CeScO_3$ can be also prepared by arc-melting of $CeO_2$, $Sc_2O_3$, and Sc metal.[22] This compound has also interesting magnetic properties.[22]

The crystal structure of $CeScO_3$ is shown in Fig. 1. Sc atoms are coordinated by six oxygen atoms forming $ScO_6$ octahedra, which are connected by sharing corners. The empty spaces formed by this arrangement are occupied by $Ce^{3+}$ ions, which are coordinated to twelve oxygen atoms. It is noticeable that the $ScO_6$ octahedron in the rare-earth scandates is more symmetric compared to the $FeO_6$ octahedron in rare-earth orthoferrites, and thus the distortion induced by pressure in the octahedral units may be different. Consequently, scandates are expected to have distinctive elastic properties,[21] though they have not explored under pressure yet. In addition, in the particular case of $CeScO_3$, a possible valence fluctuation and disproportionation can also be expected under pressure. On the other hand, the study of the HP structural properties of cerium containing oxides is extremely important for nuclear technology because they can be used as surrogate materials to simulate the thermo-physical properties of radioactive plutonium containing oxides.[23]

Since HP information on $CeScO_3$ is relevant in several areas of materials research and it is necessary to better understand its mechanical properties, we have decided to carry out by the first time a HP study on a rare-earth scandate. In particular, we have carried out HP x-ray powder diffraction (XRD) experiments on $CeScO_3$. The perovskite-type structure is found to remain stable up to 40 GPa. Its axial compresibilities and pressure-volume equation of state (EOS) have been obtained. In addition, we obtained information on bond compressibilities. Compression mechanisms and the influence of octahedral distortion and tilting in the reported results are discussed. The pressure behavior of the crystal structure of $CeScO_3$ is compared with other perovskite-oxides and cerium containing oxides.

## II. Experimental methods

CeScO$_3$ was synthesized by a two-step synthesis route involving a combustion method followed by vacuum heating at 1100 °C in the presence of a Zr sponge which acts as an oxygen getter.[21] Analysis of the final product by XRD, scanning electron microscopy, and chemical analyses confirmed that the synthesized samples consist of single phase perovskite-type CeScO$_3$. Three series of HP powder XRD experiments were performed. The maximum pressure reached was 40 GPa. Angle-dispersive XRD experiments were carried out using diamond-anvil cells (DAC) with diamond culets of 300 μm. The pressure chamber was a 100-μm hole drilled on a 40-μm pre-indented tungsten gasket. We used 16:3:1 methanol-ethanol-water as pressure-transmitting medium (PTM). Pressure was determined always by using two methods, Cu as internal pressure standard[24] and ruby fluorescence.[25] The error in the pressure determination was 0.05 GPa for pressure below 10 GPa and 0.1 GPa for the rest of the pressures. Special care was taken to occupy only a small fraction on the pressure chamber with sample and pressure standard to reduce the possibility of sample bridging between the two diamond anvils.[26] XRD experiments were performed at the materials science and powder diffraction (MSPD) beamline of the ALBA synchrotron facility.[27] The beamline is equipped with Kirkpatrick-Baez mirrors to focus the monochromatic beam to 20 μm x 20 μm (Full width at half maximum) and a Rayonix CCD detector with a 165 mm diameter of active area. We used a monochromatic wavelength of 0.4246 Å and the sample-detector distance was set to 280 mm. The two-dimmensional diffraction images were integrated with FIT2D software.[28] Structural analyses were performed with PowderCell[29] and GSAS.[30] Calculations of bond distances and polyhedral volumes were carried out using Vesta.[31]

## III. Results and Discussion

Synchrotron powder XRD experiments diffraction confirmed the perovskite-type (Pnma) structure of $CeScO_3$ at the lowest pressure achieved in this study (0.2 GPa). A Rietveld refinement of the diffraction pattern collected at 0.2 GPa gives the following unit-cell parameters: $a$ = 5.777(6) Å, $b$ = 8.045(9) Å, and $c$ = 5.642(6) Å. The structure has four formula units per unit cell (Z = 4) and the unit-cell volume is 262.2(8) Å$^3$. The Rietveld refinement of the XRD profile measured at 0.2 GPa, assuming the orthorhombic perovskite structure (Pnma), converged to small R-factors: $R_p$ = 2.67 %, $R_F^2$ = 2.28 %, and $R_{WP}$ = 3.70 % (59 Bragg reflections). In Fig. 2 (bottom part), we show the good agreement between the calculated profile and the experiment. The theoretical profile of the peaks is calculated using a pseudo-Voigt function. In addition to the unit-cell parameters, cerium and oxygen positions were refined (Sc positions are fixed by the structure symmetry). Since O has a smaller x-ray scattering cross section than Ce, it is difficult to accurately obtain simultaneously the two atomic coordinates of Ce and the five atomic coordinates corresponding to the two different oxygen atoms. Therefore, first the Ce position was determined by fixing the oxygen atoms at the ambient pressure positions. Subsequently the Ce atom was fixed and the atomic positions of the oxygen atoms were determined. This procedure was successfully used in the past to determine oxygen positions from HP experiments.[32, 33] In the case of $CeScO_3$, it conducts to consistent results which allowed to calculate bond distances and other magnitudes like the tilting angle up to 12.7 GPa. The errors for the oxygen coordinates are estimated to be approximately two times the error of Ce coordinates (see Table 1). In the refinement of the XRD data collected in the DAC, the isotropic displacement factors (B) were fixed at those values determined at ambient pressure,[21] which are mainly dominated by the background of the XRD pattern. Often the atomic displacement factors are correlated and more sensitive to background subtraction than positional parameters. The atomic positions, obtained for the structure, are summarized in Table 1. The obtained structural parameters are similar to those determined at ambient pressure.[21]

As shown in Fig. 1, the structure of $CeScO_3$ can also be explained by an alternative description using the cations sub-lattice where the oxides can be related to the structure of the corresponding elements and alloys where cations are in an oxygen matrix.[34-36] In $CeScO_3$, the cations subarray adopts a distorted CsCl-type structure (with Ce and Sc replacing Cs and Cl, respectively). In this alternative cation-array view[34-36], the oxides can be represented as a pressurized version of the metallic subarray. This hypothesis is supported by the fact that Ce-Sc distances are slightly smaller in $CeScO_3$ than in cerium-scandium alloys.[37] Therefore the structural stability observed in the present study on $CeScO_3$ suggest the existence of a CeSc alloy with the CsCl structure extremely stable under compression as recently found in other metallic alloys.[38]

In order to grasp the variation of structural changes with pressure, the XRD patterns recorded at different pressures were compared. Fig. 3 shows a selection of powder diffraction patterns measured under compression. The only changes we observed in the XRD patterns are a slight shift of Bragg peaks toward high 2θ angles due to the contraction of lattice parameters. Also the shifts of peaks corresponding to the $CeScO_3$ lattice are noticeably smaller than those of Cu peaks. This suggests that $CeScO_3$ is less compressible than Cu. A typical broadening of peaks is observed in the XRD patterns under compression suggesting that non-hydrostatic conditions[39-41] were created beyond a critical pressure. For instance, the peak broadening can be noticed in Fig. 2 by comparing the XRD patterns measured at 29.1 GPa and 5.05 GPa. Though the quasi-hydrostatic limit of our pressure medium is approximately 13 GPa,[42] due to the careful DAC loading, the peak broadening is clearly detected in our experiments only beyond 23.7 GPa (see Fig. 3). No evidence of pressure-induced phase transition or chemical decomposition of $CeScO_3$ is found up to 40.4 GPa and the ambient-pressure diffraction pattern is fully recovered upon decompression (see Fig. 3). An interesting fact to remark is that different Bragg reflections evolve in a different way under compression causing successive peak splitting and merging as pressure increases. This is illustrated in Fig. 2 by the (101) and (020) peaks located near $2\theta = 6°$ which split under compression and by the (220) and (022) peaks which merge into one peak under compression. This fact is a consequence of the anisotropic compression of $CeScO_3$ and it is described below.

Rietveld refinements were successfully performed up to 12.7 GPa, with a similar quality than in the refinement of the 0.2 GPa experiment. The structural parameters obtained at 12.7 GPa are summarized in Table 1. The refinement is shown in Fig. 2 (central pattern). At this pressure the obtained R-factors were $R_p = 2.91$ %, $R_F^2 = 2.48$ %, and $R_{WP} = 3.91$ % (55 Bragg reflections). From this set of experiments (0.2 GPa ≤ P ≤ 12.7 GPa) we extracted the pressure evolution of atomic positions which are shown in Fig. 4. All atomic coordinates are affected by pressure, being the changes more prominent in the x coordinate of Ce and the z coordinate of both oxygen atoms. These changes will have consequences on the pressure evolution of bond lengths. As we described above, with the used PTM, at pressures higher than 12.7 GPa, deviatoric stresses becomes to be not negligible.[41] Therefore, no Rietveld refinement were carried out with the XRD collected beyond this pressure, but they were analyzed by using the LeBail method to extract the unit-cell parameters.[43] An example of the LeBail fits carried out at different pressures is shown in Fig. 2 (top part). The goodness-of-fit parameters of the fit shown in the figure are $R_p = 1.97$ %, $R_F^2 = 1.68$ %, and $R_{WP} = 2.87$ % (51 Bragg reflections). Our analysis confirms the orthorhombic perovskite structure for $CeScO_3$ up to 40.4 GPa.

The unit-cell parameters and volume extracted from the Rietveld and LeBail analyses of the XRD patterns are depicted in Figs. 5 and 6, respectively. The variation of the unit-cell parameters with pressure confirms the relatively incompressible nature of CeScO$_3$. The change in the volume from ambient pressure to 40 GPa is only 15%. On the other hand, we found that there is a small anisotropy on the axial compressibility. In Fig. 4, it can be seen that apparently *b* is the most compressible axis and *c* the less compressible axis in the range of 0 to 40 GPa. In particular, from our experiments we obtained at ambient pressure the linear axial compressibilities: $k_a = \frac{-1}{a}\frac{da}{dP} = 2.10\ 10^{-3}$ GPa$^{-1}$, $k_b = \frac{-1}{b}\frac{db}{dP} = 2.35\ 10^{-3}$ GPa$^{-1}$, and $k_c = \frac{-1}{c}\frac{dc}{dP} = 1.70\ 10^{-3}$ GPa$^{-1}$. These values were determined after a Murnaghan EOS fit to experimental data.[44] The fits are shown as solid lines in Fig. 5. A consequence of the different axial compressibilities is the distinctive pressure evolution of Bragg peaks observed in Fig. 3. In particular, peaks like (020) should shift to higher angles more quickly than peaks like (101), as we commented above. Another consequence of the anisotropic compressibility is that *a* gradually approaches *c*, becoming the difference between them smaller than 1% at approximately 35 GPa. As a result of it, the (220) and (022) peaks gradually merge under compression as we described above. However, an extrapolation of the pressure dependence of both axes (*a* and *c*) to higher pressures, using the Murnaghan fits shown in Fig. 5, indicates that *a* and *c* will only become identical beyond 60 GPa. Therefore, a possible second-order orthorhombic-tetragonal transition is expected to take place only under extreme compression. A similar reduction of the lattice distortion has been observed in other materials like InSe; which transforms from monoclinic to tetragonal under HP, going through a second-order transition.[45] An extension of the pressure limits of this study will be needed to search for the existence of a pressure-driven phase transition in CeScO$_3$ and to elucidate if such transition will be a second-order orthorhombic-tetragonal transition or a first-order one.

The anisotropy in the axial compressibility of CeScO$_3$ ($\beta_a : \beta_b : \beta_c$ = 1.23 : 1.38 : 1.00; being $\beta_x = k_x/k_c$) is similar to that of other orthorhombic perovskites ($b \geq a \geq c$), like GdAlO$_3$ ($\beta_a : \beta_b : \beta_c$ = 1.24 : 1.65 : 1.00)[12], LaGaO$_3$ ($\beta_a : \beta_b : \beta_c$ = 1.61 : 1.68 : 1.00)[13], while the anisotropy in compression is smaller in the case of CaTiO$_3$ ($\beta_a : \beta_b : \beta_c$ = 1.05 : 1.02 : 1.00)[46] and GdFeO$_3$ ($\beta_a : \beta_b : \beta_c$ = 1.03 : 1.06 : 1.00)[12]. Thus, orthorhombic structures like GdAlO$_3$ and CeScO$_3$ are likely to evolve under HP towards higher symmetry structures. From *ab initio* calculations, Wu *et al.* have predicted for CaTiO$_3$ a structural transition from Pbnm (equivalent to Pnma) to Cmcm at pressures above 40 GPa.[47] However, the experiments of Zhao *et al.* suggested that the structure would likely to transform to a more distorted structure due to the differences in the compressibilities of the CaO$_{12}$ and TiO$_6$ polyhedral units.[48] This aspect is explained later in this article. Since the TiO$_6$ octahedron is less compressible, it will tilt more and thus the orthorhombic CaTiO$_3$ will transform to a more distorted structure while approaching a ~ c.

The present pressure-volume data have been analyzed using a third-order Birch-Murnaghan (BM) EOS[44] by the EOSFit7 software.[49] If the data obtained from the complete pressure range covered by experiments are included in the fit, the obtained volume at ambient pressure (V$_0$), bulk modulus (B$_0$), and its pressure derivative (B$_0$') are V$_0$ = 262.5(3) Å$^3$, B$_0$ = 165(7) GPa, and B$_0$' = 6.3(5), respectively. The goodness of fit ($\chi^2$) is 1.45. According with the EOS parameters determined, the implied value[50] for the second pressure derivative is B$_0$'' = -0.0696 GPa$^{-1}$. If only data obtained for P ≤ 12.7 GPa are included, we got V$_0$ = 262.3(3) Å$^3$, B$_0$ = 177(3) GPa, and B$_0$' = 4.0(6) with $\chi^2$ = 1.15; i.e. a second-order BM EOS describes well the results obtained under quasi-hydrostatic conditions (P ≤ 12.7 GPa). The differences in the bulk modulus and its derivative (which are correlated) between both fits reflect the change of compressibility observed in the experiments near 10 GPa. The bulk modulus obtained for perovskite-type CeScO$_3$ (B$_0$ =

165 – 177 GPa) indicates that this oxide is less compressible than zircon-type CeVO$_4$ (B$_0$ = 133 GPa)[51] and monazite-type CePO$_4$ (B$_0$ = 110 GPa)[52], but more compressible than fluorite-type CeO$_2$ (B$_0$ = 220 GPa)[53] and Ce$_2$Zr$_2$O$_8$ (214 GPa)[23]. An explanation to this observation comes from two facts: a) in all compounds mentioned above most of the crystal compressibility is expected to come from the contraction of Ce-O bonds due to their softer nature, and b) also due to the symmetry and distortion in the local coordination, as they account for the bond valence sums at the cation sites. A relatively lower compressibility is expected in CeO$_2$ and Ce$_2$Zr$_2$O$_8$ as they have Ce$^{4+}$-O$^{2-}$ bonds, which are shorter and stronger compared to Ce$^{3+}$-O$^{2-}$ bonds in CeVO$_4$, CePO$_4$, and CeScO$_3$. The local symmetry and distortions are the plausible reason for the larger compression in these structures. Further conclusions on the mechanism of compressibilities are obtained by comparing the local structural parameters and effect of pressure on the CeO$_{12}$ and ScO$_6$ coordination polyhedra.

In general, the tilt patterns and distortion of the octahedral units of perovskite-type materials depend on the ionic radii of the cations as well as external parameters like temperature and pressure, and they govern the structure and phase transitions. Usually the octahedral tilt is reduced with increasing temperature while an opposite trend is observed with pressure. Thus, the symmetry of the perovskite lattice increases with increasing temperature while it decreases with increasing pressure. However, the general observation of pressure-induced structural transition is not the straightforward inverse relation as mentioned earlier. The "anomalous" behavior of pressure-induced structural transition has been attributed to the distortion and relative compressibility of the polyhedra around the cations. In the case of CeScO$_3$, the experimental observations indicate the absence of any structural transition up to 40 GPa.

It has been observed that the bulk modulus of various oxides can be explained in terms of local compressibilities.[54-58] In order to further understand the structural evolution and to compare with the general trend of phase transition the compressibility of Ce-O and Sc-O bonds is investigated. In addition, the pressure evolution of polyhedral distortions and polyhedral compressibility were analyzed to understand the microscopic effect of pressure. We have done it for the pressure range where Rietveld refinements were performed (P ≤ 12.7 GPa). The obtained results on bond compressibility are summarized in Fig. 7. It can be seen from the figure that one of the Sc-$O_2$ bonds compresses more than the other Sc-$O_2$ bond and the Sc-$O_1$ bond. However, the deformation of the $ScO_6$ octahedron upon compression is small. In particular, the bond length distortion, which is defined as the relative deviation from average bond length, $\Delta_{Oct} = \frac{1}{6}\sum_{i=1}^{6}(X_i - \bar{X})^2/\bar{X}^2$,[59,60] changes from $9.8 \times 10^{-5}$ at 0.2 GPa to $6.0 \times 10^{-5}$ at 12.7 GPa. If the Robinson parameters[61] are used to evaluate the pressure influence in the $ScO_6$ octahedron, the changes in its shape from ambient pressure up to 12.7 GPa are not detectable, being the quadratic elongation nearly constant and close to 1.0006. This behavior is qualitatively similar to that obtained from single-crystal XRD for the $AlO_6$ and $SnO_6$ octahedron in $YAlO_3$[16] and $CaSnO_3$[17], respectively. Thus, with increasing pressure, the $ScO_6$ octahedron becomes more symmetric, and its volume sphericity increases from 0.967 to 0.975 as pressure increases from 0.2 GPa to 12.7 GPa. A similar analysis of the distortion in the $CeO_{12}$ polyhedral units indicates only a marginal decrease in bond length distortion (from $2.41 \times 10^{-2}$ to $1.51 \times 10^{-2}$ in the same pressure range). Thus the $CeO_{12}$ unit becomes slightly more symmetric under compression. On the other hand, the long Ce-O bonds are in average 25 % more compressible than the Sc-O bonds. The difference found in bond compressibilities gives for the $ScO_6$ octahedron a bulk modulus of 197(2) GPa and for the $CeO_{12}$ units a bulk modulus of 174(3) GPa. The polyhedral volumes of $CeO_{12}$ and $ScO_6$ calculated from the structural data at different pressures are shown in Fig. 8, which indicates a decreasing trend in both while the decease is more prominent in the $CeO_{12}$ cubooctahedra compared to the $ScO_6$ octahedra. This result is in agreement with the greater bulk modulus for $ScO_6$ than for $CeO_{12}$, as reported above. Thus, the compressibility of $CeScO_3$ perovskite ($B_0$ = 177 GPa when same pressure range is evaluated) is closely similar to that of $CeO_{12}$ cuboctahedron. This observation is consistent with the polyhedral compressibility model, developed by Recio *et al.* for spinel oxides,[54] which predicts the $CeO_{12}$ cubooctahedra (whose volume at 0.2 GPa is 4.45 times larger than the volume of the $ScO_6$ octahedra) to dominate the compressibility of $CeScO_3$. A similar behavior has been observed in $ScAlO_3$ and other perovskites.[6] The polyhedral volume ratio of $CeO_{12}$ and $ScO_6$ ($V_A/V_B$) at 0.2 GPa, 4.45, decreases only marginally until 4.43 at 12.7 GPa.

If the octahedral units are so rigid, the only way in which the ionic packing can be more efficient is by introducing tilts of these units. Therefore, pressure will result in an increase of tilting without a change of symmetry. From the Rietveld refinements carried out up to 12.7 GPa we have obtained information on the influence of pressure in the tilting angles which is shown in Fig. 9. This information is relevant since many physical properties as well as structural transitions of perovskites are derived from the degree of octahedral tilts.[62] The observed variation of the tilting angles is consistent with a rigid octahedron model. For the ideal cubic perovskite structure, these angles are 180° and a deviation of these angles from 180º increases the lattice distortion while lowering the symmetry as well as the unit-cell volume. The behavior of the tilting angles of $CeScO_3$ resembles the results reported from $CaSnO_3$ and $CaTiO_3$[17, 46], but it is different from the behavior reported for magnetic $GdMnO_3$[11] and $YAl_{0.25}Cr_{0.75}O_3$.[63] On the other hand, in Fig. 9 it can be seen that the large tilting angle (Sc-$O_2$-Sc ~ 148.7° at 0.2 GPa) decreases more rapidly than the small tilting angle (Sc-$O_1$-Sc). From the structure it can be visualized that the Sc-$O_2$-Sc connections are propagating along the <101> direction and thus the orthorhombic crystal structure evolves towards a higher symmetry, which is coherent with the fact that the *a*- and *c*-axes gradually approach the same value under compression.

Before concluding, we would like to comment on the pressure stability of $CeScO_3$. Several orthorhombic perovskites are known to undergo phase transitions at relative low pressure. For instance, $LaGaO_3$ experience an orthorhombic to rhombohedral phase transition near 2.5 GPa,[7] $CaZrO_3$ goes through an orthorhombic to monoclinic phase transition below 30 GPa,[64] and $NaMgF_3$ undergoes a phase transition at 27 GPa.[3] After an exhaustive survey of a large number of perovskite-type structures, Martin and Parise[3] have established a phenomenological criterion, based upon the ratio between polyhedral volumes, to estimate transition pressure in perovskites. In particular, these authors propose that a perovskite becomes unstable when the polyhedral volume ratio of the cuboctahedron ($CeO_{12}$ in our case) and the octahedron ($ScO_6$ in our case), $V_A/V_B$, is smaller than four. In our case, this coefficient is 4.45 at 0.2 GPa, decreasing under compression and reaching a value of 4.42 at 12.7 GPa, as previously commented. A linear extrapolation to higher pressure indicates that at 40 GPa this parameter will reach 4.36, which is way far from the instability value of 4, indicating that $CeScO_3$ should be still stable, which is consistent with the fact that we did not find any structural phase transition. Similarly, Angel *et al.*[5] have suggested that that the ratio between polyhedral compressibilities $k_A/k_B$ guides the distortion of perovskites under pressure; viz. for $k_A/k_B = 1$ the structure remains unchanged while for $k_A/k_B > 1$, the structure transforms to a lower symmetry structure and for $k_A/k_B < 1$, the structural distortion increases and the structure transforms to a higher symmetry structure. For $CeScO_3$, $k_A/k_B$ is 1.09 (~1) and thus the structural transition is not expected according to all other observations. Due to a similar reason ($k_A/k_B \sim 1$), the cubic BaLiF$_3$ and KMF$_3$ compounds also do not undergo any pressure induced structural transition.[9, 10]

Finally, the HP stability of CeScO$_3$ can be also understood in terms of its cation network, which we have previously described; i.e. based upon the packing of Ce and Sc atoms. Since the ionic radius of Sc$^{3+}$ (0.745 Å) is small relative to that of Ce$^{3+}$ (1.34 Å), in CeScO$_3$, the stresses induced by compression can be accommodated by deformation of the Ce outer shell as opposed to significant changes in its atomic position, thereby favoring the stability of the perovskite structure under compression.[65] In contrast, perovskites like CeSrO$_3$, in which Ce$^{4+}$ (1.14 Å) and Sr$^{2+}$ (1.18 Å) have comparable ionic radii, should accommodate compression by displacements of Ce and Sr from their positions resulting in a pressure-induced phase transition, which has been estimated to occur at 22 GPa.[3] Following this reasoning, other alkaline-earth scandates are expected to be extremely stable under compression as we observed in CeScO$_3$. Moreover, the perovskites containing transition metal ions like Mn$^{3+}$, Ru$^{4+}$, Co$^{3+}$, etc. show transitions with volume discontinuities due an electronic structural transition.[4, 9, 11]

**IV.    Concluding Remarks**

In this work, we reported an experimental study of the structural properties of CeScO$_3$ under compression. This is a first effort to understand the HP behavior of rare-earth scandates. Our synchrotron powder x-ray diffraction experiments have allowed us to determine that the initial phase remains stable in the orthorhombic perovskite structure up to very high pressure (at least up to the highest pressure covered by experiments, 40 GPa). The pressure-volume room-temperature equation of state of this phase is determined together with the pressure dependence of the unit-cell parameters. In particular, a bulk modulus of 165 GPa is determined for perovskite-type $CeScO_3$. Information on bond compressibility and tilting angles is finally presented. The reported results are compared with related perovskites and with other ternary oxides containing cerium atoms. Reasons for the observed compressibility and the pressure stability of $CeScO_3$ are given. We hope this first high-pressure study on rare-earth scandates will trigger further studies on this research topic.

**AUTHOR INFORMATION**

**Corresponding Author**

*E-mail: daniel.errandonea@uv.es.

**Author Contributions**

The manuscript was written through contributions of all authors. All authors have given approval to the final version of the manuscript.

**Notes**

The authors declare no competing financial interests.

**Acknowledgments**

The authors thank the financial support to this research from the Spanish Ministerio de Economía y Competitividad (MINECO), the Spanish Research Agency (AEI), and the European Fund for Regional Development (FEDER) under Grants No. MAT2016-75586-C4-1/2-P, MAT2013-46649-C4-1/2-P, and No. MAT2015-71070-REDC (MALTA Consolider). D.S.P. acknowledges the Spanish government for a Ramon y Cajal grant. The authors express gratitude to F. Aguado for fruitful discussions on the high-pressure behavior of perovskites. These experiments were performed at MSPD beamline at ALBA Synchrotron with the collaboration of ALBA staff.

**Table Captions**

**Table 1:** Structural parameters at 0.2 (top) and 12.7 (bottom) GPa; space group: Pnma. The assumed values for the isotropic displacement parameters B are also given. The unit-cell parameters determined are $a = 5.777(6)$ Å, $b = 8.045(9)$ Å, $c = 5.642(6)$ Å (at 0.2 GPa) and $a = 5.639(7)$ Å, $b = 7.897(9)$ Å, $c = 5.533(7)$ Å (at 12.7 GPa).

**Figure Captions**

**Figure 1:** Schematic view of the crystal structure of $CeScO_3$. $ScO_6$ octahedral units are shown in green, the oxygen atoms in red ($O_1$) and orange ($O_2$), and the Ce atoms in gray. One $CeO_{12}$ polyhedron is shown in right-hand side of the figure.

**Figure 2:** Selection of Rietveld (0.2 and 11.6 GPa) and LeBail (29.1 GPa) refinements of XRD patterns. Experiments are shown as dots, refinements and residuals as solid lines. The ticks indicate the position of the Bragg reflections for Cu and $CeScO_3$.

**Figure 3:** Selection of HP XRD patterns. Pressures are indicated in the plot. The most intense reflection from Cu, corresponding to the (111) Bragg peak, which is used to determine pressure, is labeled in all diffraction patterns. A pattern collected after decompression is shown denoted by (r). The indexes of the reflections discussed in the text are shown.

**Figure 4:** Pressure dependence of free atomic coordinates up to 12.7 GPa. Symbols: refined positions. The size of the symbols is comparable to error bars. The lines are a guide to the eyes.

**Figure 5:** Pressure dependence of unit-cell parameters. Different symbols correspond to the three different experiments. The size of the symbols is comparable to error bars. The lines are a guide to the eyes. Lines represent the fits described in the text.

**Figure 6:** Unit-cell volume versus pressure. Different symbols correspond to the three different experiments. The size of the symbols is comparable to error bars. The lines are a guide to the eyes. The solid line shows the third-order Birch-Murnaghan EOS obtained considering the whole pressure range covered by experiments.

**Figure 7:** Bond lengths versus pressure. Symbols: experimental results. Error bar are shown when larger than symbol size. Lines: linear fits.

**Figure 8:** Polyhedral volume of $CeO_{12}$ and $ScO_6$ as a function of pressure. The size of the symbols is comparable to error bars. The lines are a guide to the eyes.

**Figure 9:** Tilting angles as function of pressure. Symbols are the angles calculated from the refined structures. The solid lines are linear fits to the data to illustrate the pressure dependence of the tilting angles.

**Table 1**

| Atom | Wyckoff position | x | y | z | B [Å$^2$] |
|---|---|---|---|---|---|
| Ce | 4c | 0.0375(5) | 0.25 | 0.9808(5) | 0.250 |
| Sc | 4b | 0 | 0 | 0.5 | 0.732 |
| O$_1$ | 4c | 0.977(1) | 0.25 | 0.386(1) | 1.112 |
| O$_2$ | 8d | 0.298(1) | 0.053(1) | 0.706(1) | 0.863 |

| Atom | Wyckoff position | x | y | Z | B [Å$^2$] |
|---|---|---|---|---|---|
| Ce | 4c | 0.0392(5) | 0.25 | 0.9798(5) | 0.250 |
| Sc | 4b | 0 | 0 | 0.5 | 0.732 |
| O$_1$ | 4c | 0.977(1) | 0.25 | 0.383(1) | 1.112 |
| O$_2$ | 8d | 0.298(1) | 0.054(1) | 0.703(1) | 0.863 |

**Figure 1**

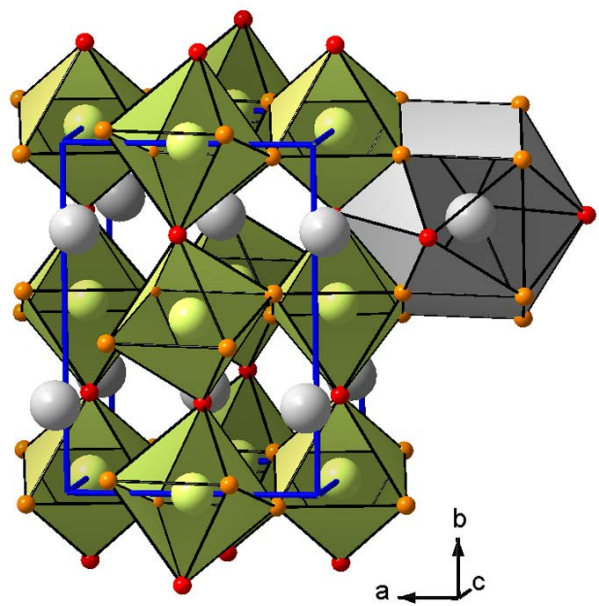

**Figure 2**

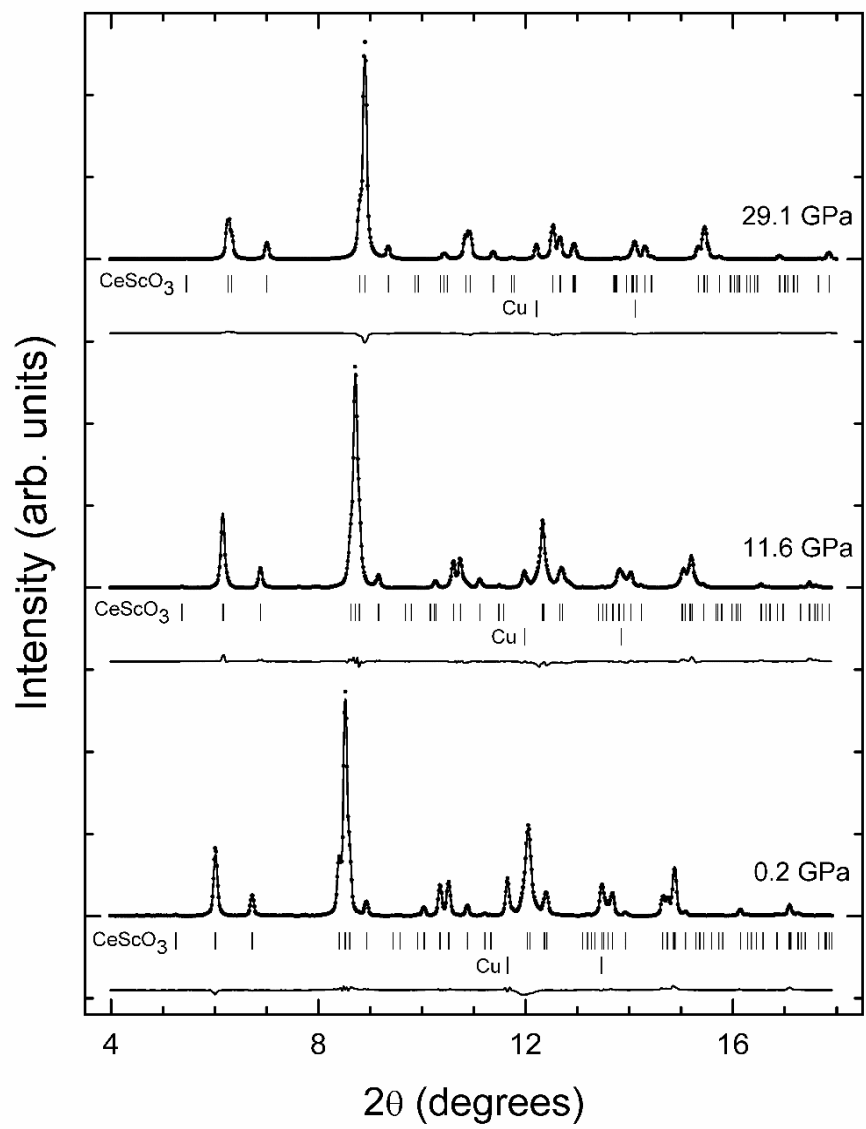

**Figure 3**

**Figure 4**

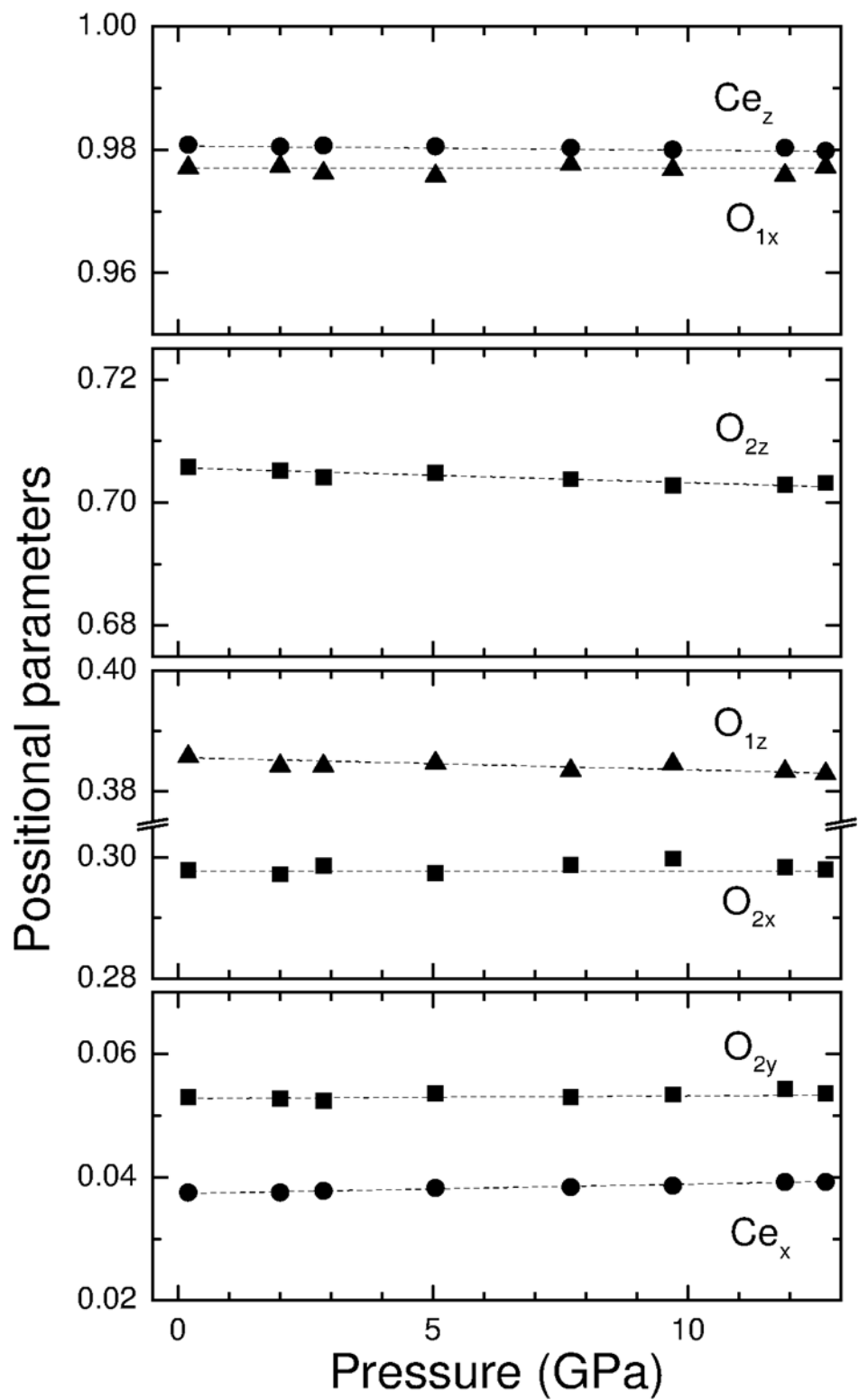

**Figure 5**

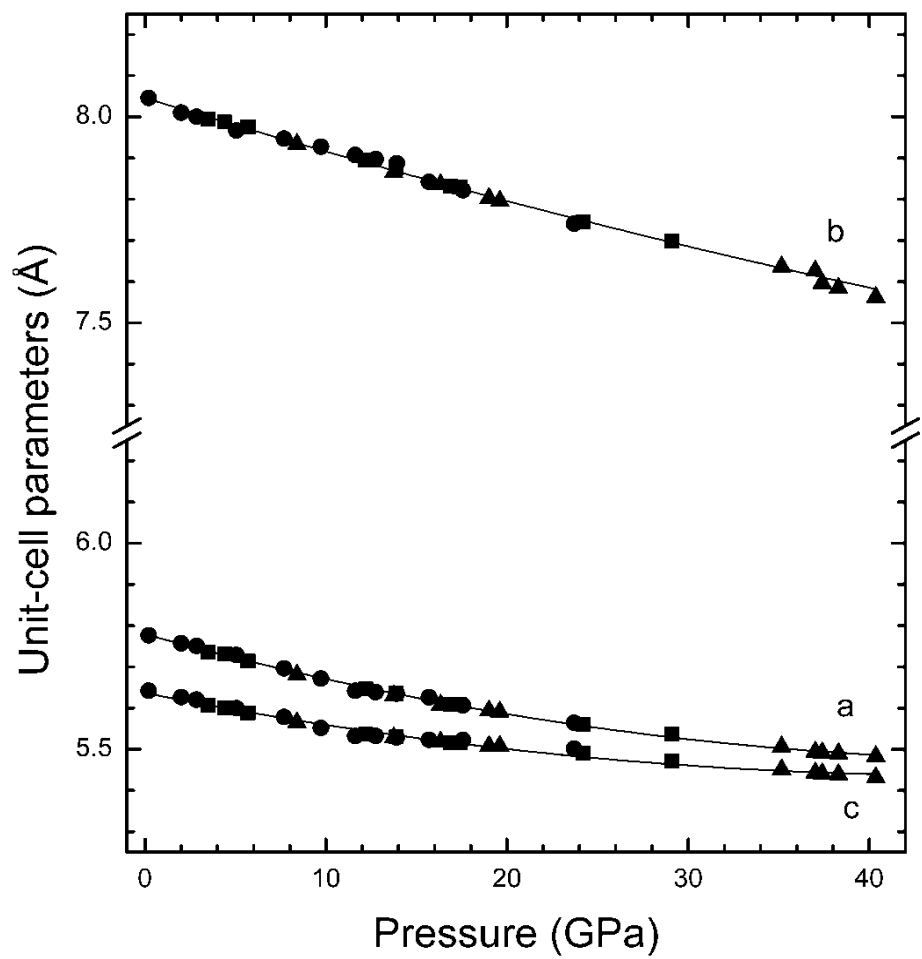

**Figure 6**

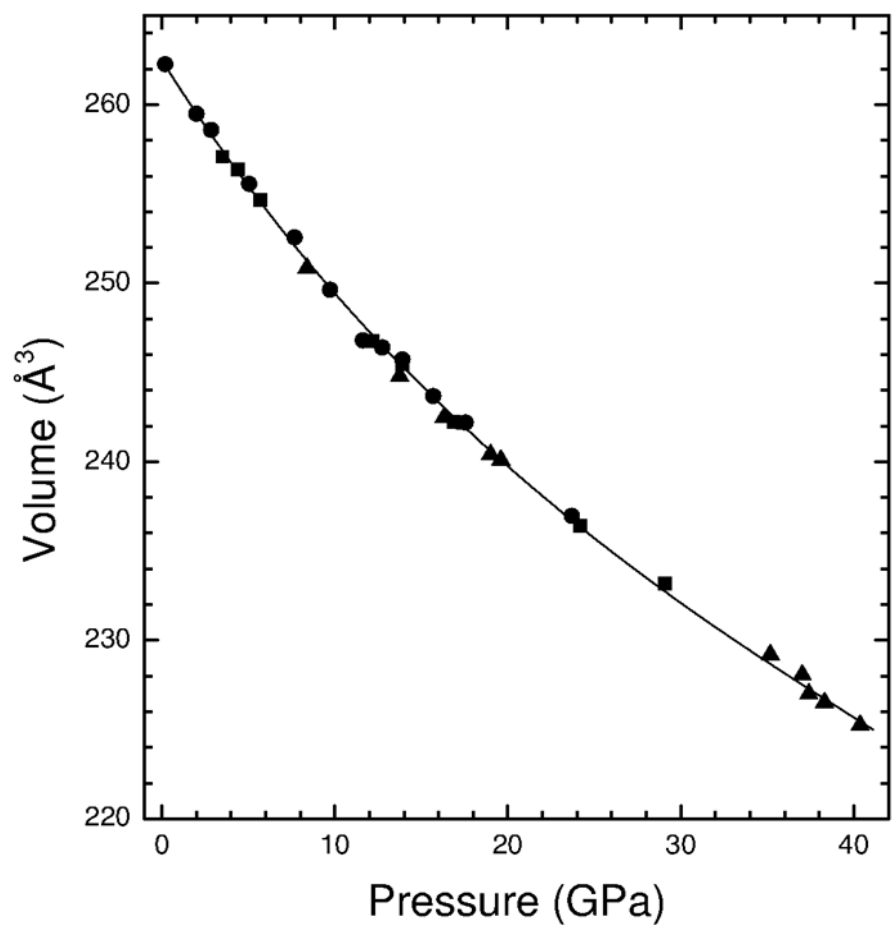

**Figure 7**

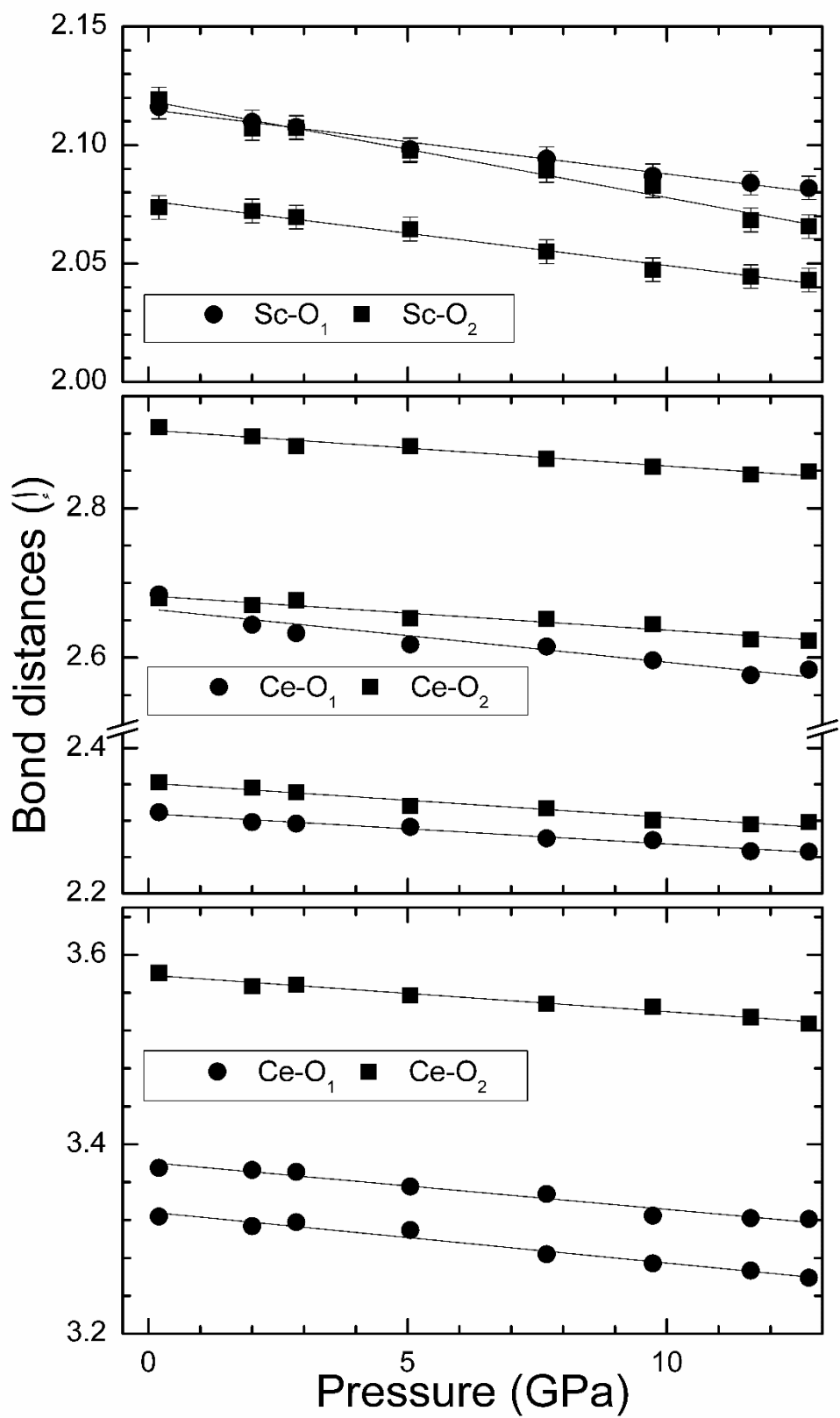

**Figure 8**

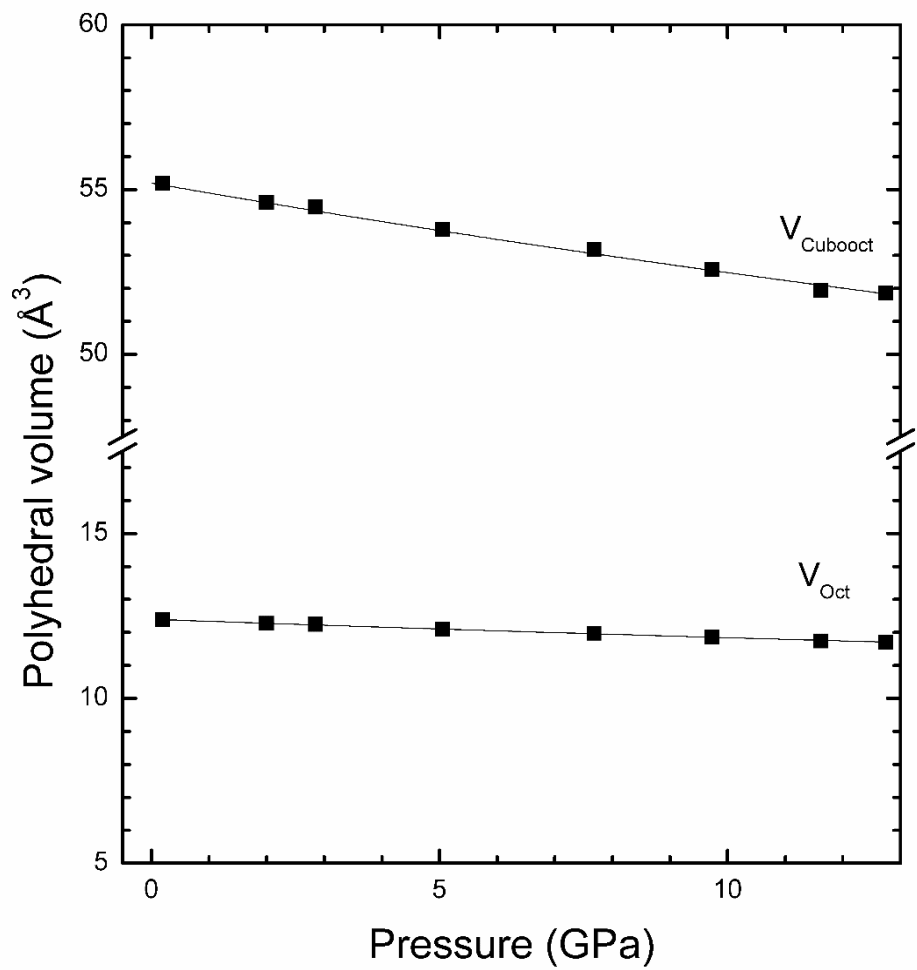

**Figure 9**

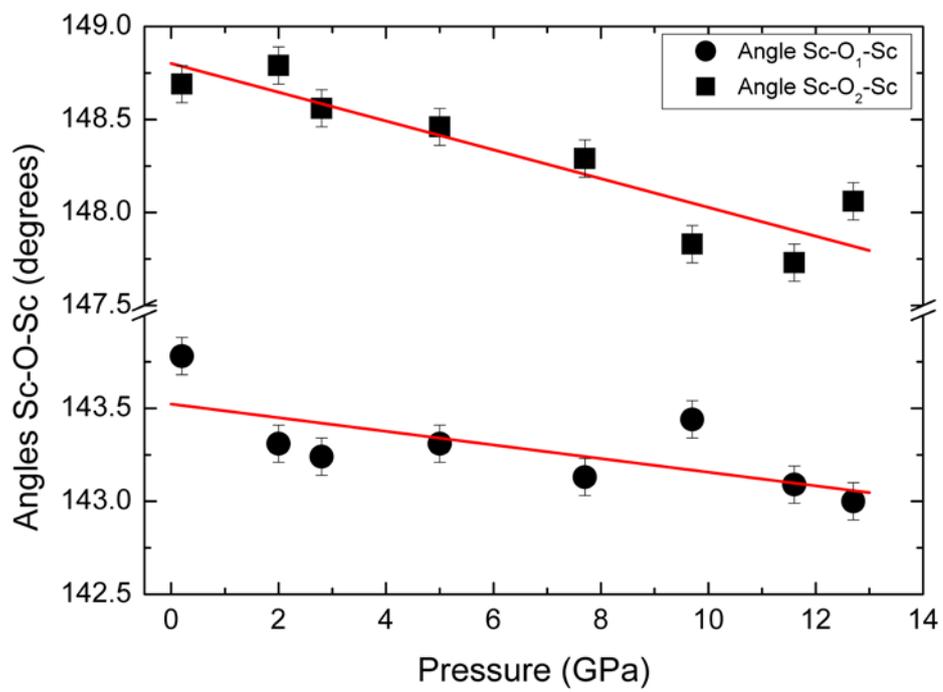

**For Table of Contents Only**

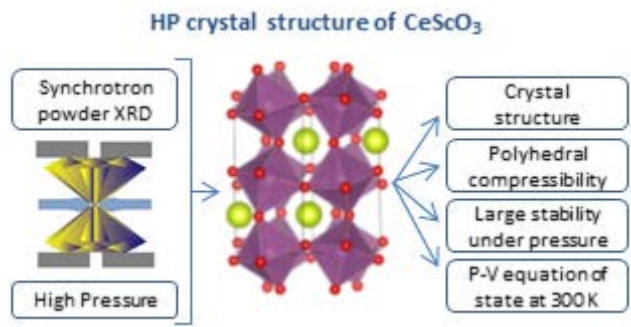

The high-pressure structural of perovskite-type $CeScO_3$ were studied for the first time up to 40 GPa. The room-temperature equation of state and axial and bond compressibilities are reported. Macroscopic changes are related to changes in bond distances and the octahedral tilt. Reasons for the large pressure stability of $CeScO_3$ are discussed.